\documentclass[a4paper,11pt,english]{article}
\usepackage[T1]{fontenc}
\usepackage{geometry}
\usepackage{amssymb,amsmath}
\usepackage{hyperref}
\usepackage{graphicx}
\usepackage{cite,mcite}
\usepackage{enumerate}
\usepackage[usenames,dvipsnames]{color}

\newcommand{\executeiffilenewer}[3]{%
\ifnum\pdfstrcmp{\pdffilemoddate{#1}}%
{\pdffilemoddate{#2}}>0%
{\immediate\write18{#3}}\fi%
}

\global\parskip 6pt

\makeindex

\begin{document}

\begin{center}
{\Large{\bfseries Bulk quartic vertices\\ from\\ Boundary four-point correlators

}}
\vskip 0.05\textheight

{\bf Xavier Bekaert\footnote{xavier.bekaert@lmpt.univ-tours.fr}, Johanna Erdmenger\footnote{jke@mpp.mpg.de}, Dmitry Ponomarev\footnote{d.ponomarev@imperial.ac.uk} and Charlotte Sleight\footnote{csleight@mpp.mpg.de}}\\

\vskip 0.5cm
\emph{${}^{1} \:$Laboratoire de Math\'{e}matiques et Physique Th\'{e}orique}\\ 
\emph{Unit\'{e} Mixte de Recherche 7350 du CNRS}\\  
\emph{F\'{e}d\'{e}ration de Recherche 2964 Denis Poisson}\\ 
\emph{Universit\'{e} Fran\c{c}ois Rabelais, Parc de Grandmont, 37200 Tours, France}
\vskip 0.5cm
\emph{${}^{2,\;4}\:$Max-Planck-Institut f\"{u}r Physik (Werner-Heisenberg-Institut)}\\
\emph{F\"{o}hringer Ring 6, D-80805 Munich, Germany}
\vskip 0.5cm
\emph{${}^{3}\:$The Blackett Laboratory,}\\ 
\emph{Imperial College,}\\
\emph{London SW7 2AZ, United Kingdom}
\vskip 0.5cm

\end{center}


\begin{center} 
	{\bf Abstract }  
\end{center} 
\begin{quotation}
We present arguments which suggest that the bulk higher-spin gravity duals of weakly-coupled conformal field theories  obey some refined notion of locality.  In particular, we discuss the Mellin amplitude programme in this context. We focus on the $O(N)$ vector model and minimal higher-spin gravity as a paradigmatic example of such holographic dual pairs. We restrict ourselves to three- and four-point functions of scalar primary operators, but the qualitative conclusions are expected to hold for the generic case.
\end{quotation}

\newpage
\restoregeometry

\pagestyle{plain}

\tableofcontents

\vspace{3cm}


\section{Bulk locality and Mellin amplitudes}\label{bulkloc}
Bulk locality is one of the most remarkable features of the AdS/CFT correspondence. The concept of bulk locality refers to the fact that the usual conditions for locality in a quantum field theory also hold in the interior of anti de Sitter (AdS) space. It is expected to hold in the usual regime where the duality is tested, i.e. when the curvature radius $R_{AdS}$ is large compared to the string length $\ell_s$ which, in turn, is large compared to the Planck length $\ell_P$. On the gravity side, this corresponds to
(I) a semi-classical limit: $R_{AdS}\gg \ell_P$ and (II) a higher-spin gap: all particles of spins $s$ greater than two have large masses $M_{s>2}\gg M_{s'\leq 2}$.
On the conformal field theory (CFT) side, this translates into: 
\begin{itemize}
		\item[(I)] a semi-classical limit: a perturbative expansion around a generalised free field theory when $N\gg 1$,
		\item[(II)] a higher-spin gap: all single-trace operators of higher-spins have large conformal dimensions $\Delta_{s>2}\gg \Delta_{s'\leq 2}$\,.
\end{itemize}
These two properties were argued to provide necessary and sufficient conditions for a CFT to possess a semiclassical and local bulk dual \cite{Heemskerk:2009pn,Fitzpatrick:2010zm}. Later, a third condition was added from the perspective of effective field theory \cite{Fitzpatrick:2012cg}:
\begin{itemize}
		\item[(III)] polynomial boundedness of Mellin amplitudes.
\end{itemize}

Over the last years, significant progress has been made in the Mellin amplitude programme. Initiated in \cite{Penedones:2010ue}, the goal is to interpret the Mellin amplitudes \cite{Mack:2009mi} associated to CFT correlation functions as AdS scattering amplitudes
(the corresponding dictionary is summarised in Table \ref{Mampldict}).

\begin{table}[h]
 \centering
\caption{Mellin amplitude dictionary}
{\begin{tabular}{|c|c|}
 \hline
Minkowski spacetime & Anti-de Sitter spacetime \\\hline\hline
Feynman diagram & Witten diagram \\\hline
Mandelstam invariants & Mellin variables \\\hline
Scattering amplitude & Mellin amplitude \\\hline
\end{tabular}
}
\label{Mampldict}
\end{table}

For a given correlation function, the corresponding Mellin amplitude is a suitably normalised Mellin transform of the factor that is not fixed by conformal symmetry. Consider, for example, the simplest case of a four-point correlator of a scalar primary operator ${\cal O}$ with scaling dimension $\Delta$,
\begin{equation}
\langle \mathcal{O}\left(y_1\right) \mathcal{O}\left(y_2\right) \mathcal{O}\left(y_3\right) \mathcal{O}\left(y_4\right)\rangle
= \frac{1}{\left(y^2_{12} y^{2}_{34}\right)^{\Delta}} \,F(u,v)\,,
\label{4pt}
\end{equation}
with $u$ and $v$ the two cross ratios $u = \frac{y^2_{12}y^2_{34}}{y^2_{13}y^2_{24}}$, $v = \frac{y^2_{14}y^2_{23}}{y^2_{13}y^2_{24}}$, and $y_{ij} = y_i - y_j$. Then the \textit{reduced Mellin amplitude} \cite{Mack:2009mi} 
is the function $M(s,t)$ of two scattering variables $s$ and $t$ defined via the relation
\begin{equation}
F(u,v)= \int\frac{ds}{2\pi i}\int \frac{dt}{2\pi i}\, u^s\, v^{\Delta-s-t}\, M(s,t),
\end{equation}
providing a Mellin-Barnes representation of the conformal correlator \eqref{4pt}. The \textit{Mellin amplitude} itself is the function ${\cal M}(s,t)$ defined by
\begin{equation}\label{Mst}
M(s,t)=\big[\Gamma(\Delta-s)\Gamma(\Delta-t)\Gamma(s+t-\Delta)\big]^{2}{\cal M}(s,t),
\end{equation}
where a numerical coefficient (irrelevant for the present discussion) has been dropped.
The extraction of the corresponding gamma functions is important to ensure the following property: For a large-$N$ CFT with a discrete set of primary operators, the Mellin amplitudes of single-trace primary-operator correlators are meromorphic functions with simple poles determined by the twists of the single-trace operators in the OPE \cite{Mack:2009mi,Penedones:2010ue}. In particular, the reduced Mellin amplitude  $M(s,t)$ possesses extra poles with respect to the Mellin amplitude ${\cal M}(s,t)$, due to the gamma functions in \eqref{Mst}. These extra poles originate from the double-trace operators in the OPE.

\begin{table}[h]
\centering
\caption{Amplitude properties}
{\begin{tabular}{|c|c|}
 \hline
Bulk process & Boundary amplitude \\\hline\hline
Local contact interaction & Polynomial \\\hline
Particle exchange & Simple pole \\\hline
\end{tabular}
}
\label{Scamplprop}
\end{table}

For individual (Feynman or Witten) diagrams, scattering amplitudes of elementary particles and Mellin amplitudes for single-trace primary operators thus share common properties
as functions of Mandestam invariants vs Mellin variables \cite{Penedones:2010ue,Mack:2009mi,Paulos:2011ie,Fitzpatrick:2011ia} (c.f. Table \ref{Scamplprop}) suggesting a common

\vspace{1mm}
\noindent\textbf{Boundary criterion of bulk locality}: \textit{Interactions on flat/AdS spacetime are local iff the amplitudes
of the corresponding contact Feynman/Witten diagram are polynomial functions of Mandelstam invariants / Mellin variables}.
\vspace{1mm}

In this sense, the condition (III)  can be seen as the criterion of bulk locality itself, in the same way that the condition (I) stands for the criterion of existence of a semiclassical limit. Nevertheless, the condition (II) remains of interest because the presence of a higher-spin gap sets a higher-spin symmetry breaking scale in the bulk. Even if the gravity theory may be nonlocal in the sense of being an effective field theory, the previous scale controls the low-energy expansion of the theory and the validity range of its local truncations.

The Mellin amplitude programme of rewriting CFT correlators as Witten diagrams seemingly applies to the large class of strongly-coupled CFTs obeying the criteria (I)-(III). Under these hypotheses, the corresponding bulk dual would possess a simultaneously weakly-coupled \& weakly-curved regime.
Indeed, the Mellin amplitude programme seems well adapted (but presently restricted) to the holographic reconstruction of  bulk theories (or individual scattering processes) possessing a weakly-coupled \& weakly-curved limit.
However it does not apply directly to the simplest example of CFTs: free ones (or weakly coupled ones). In fact, the Mellin transform of correlators of free CFTs is often not even well defined (see section \ref{MellinON}). The bulk duals of free CFTs are conjectured to be higher-spin gravity theories which are indeed non-local in the restricted sense of locality (see section \ref{HShol}).

What could be a mild replacement of the criteria (II) and (III), 
that could provide necessary and sufficient conditions for CFTs (including weakly-coupled ones) to possess (mildly non-local) bulk duals (including higher-spin gravity theories)?
A tentative answer could be:
\begin{itemize}
	\item[(II')] a finite number of single-trace primary operators with conformal dimension below any fixed dimension, in the $N\to\infty$ limit,
	\item[(III')] analyticity of Mellin amplitudes.
\end{itemize}
The criterion (II') is inspired from the second basic property in \cite{ElShowk:2011ag}.  In spirit, it is analogous to the second assumption of the Coleman-Mandula theorem \cite{Coleman:1967ad} since the conformal dimension of a single-trace operator translates into the energy of the corresponding elementary particle in the bulk. In contrast to flat spacetime, in AdS this condition does
not rule out a tower of massless higher spins as their dimensions grow with spin. Another motivation for criterion (II') is that bulk locality would be obscure if an infinite number of fields were relevant in a scattering process at a given energy.
The criterion (III') relaxes strict locality and replaces it with the milder requirement that the coefficients in the Taylor series expansion of the amplitude decrease fast enough in order to have an infinite radius of convergence. 
In other words, the amplitude can be approximated, for any fixed accuracy, by a polynomial of sufficiently high degree, i.e. by a local interaction of sufficiently high order. 
More precisely, the criterion (III')  amounts to the following \cite{Bekaert:2015tva}

\vspace{1mm}
\noindent\textbf{boundary criterion of weak locality:} 
\textit{Interactions on flat/AdS spacetime are weakly local iff the amplitudes 
of the corresponding contact Feynman/Witten diagram are entire functions of Mandelstam invariants / Mellin variables}. 

\section{Bulk locality and higher-spin holography}\label{HShol}

The higher-spin holographic duality arose shortly after the birth of AdS/CFT correspondence and was initially motivated by the semiclassical but stringy regime ($\ell_s\gg R_{AdS}\gg \ell_P$) in the strong version of Maldacena conjecture.
A more general picture progressively emerged in a long series of paper (see \cite{Sundborg:2000wp,
Klebanov:2002ja,Sezgin:2003pt} for some early steps) but the basic idea underlying higher-spin holography is easy to summarise.

Free (or integrable) CFTs have an infinite number of global higher (sometimes called ``hidden'') symmetries 
(including conformal symmetry). Applying Noether's theorem, one deduces that their spectrum must contain an infinite tower of traceless conserved currents with unbounded spin (including spin two). Therefore, the AdS/CFT dictionary suggests that free (or some integrable) CFTs should be dual to ``higher-spin gravity'' theories, in the sense of theories whose spectra contain an infinite tower of gauge fields with unbounded spin (including spin two). The simplest example of such a scenario states that the singlet sector of the free and critical vector models should be holographically dual to minimal higher-spin gravity \cite{Klebanov:2002ja,Sezgin:2003pt}.

Turning back to the issue of bulk locality, higher-spin interactions in four (and higher) dimension appear to be generically:
\begin{itemize}
	\item[(A)] \textbf{quasi-local} in the sense that they possess a perturbative expansion (in powers of
fields and their derivatives) where each individual term in the Lagrangian is local (effective field theories are typically quasilocal).
\vspace{1mm}
	\item[(B)] \textbf{non-local} in the sense that the total number of derivatives in the complete Lagrangian is unbounded (as in string field theory).	This is a corollary of:
		\begin{itemize}
			\item[(a)] \textit{Metsaev bounds:} The number of derivatives appearing in a non-trivial cubic vertex evaluated on the free mass-shell is bounded from  \cite{Metsaev:2005ar}
			\begin{itemize}
				\item[i.] below by the highest spin involved,
				\item[ii.] above by the sum of the spins involved.
			\end{itemize}
			\item[(b)] \textit{Higher-spin algebra structure:} The Jacobi identity requires a spectrum with an infinite tower of fields with unbounded spin \cite{Fradkin:1986ka,Boulanger:2013zza}.
		\end{itemize}
	The number of derivatives is bounded from below by the highest spin involved by virtue of Metsaev's lower bound (a.i). However, there cannot be any upper bound on the spin in a consistent theory due to point (b). Hence, the number of derivatives is necessarily unbounded.
\vspace{1mm}
	\item[(C)] \textbf{weakly-local} in the sense that the Mellin amplitude of contact Witten diagrams are entire functions of the Mellin variables.
	
\vspace{1mm}	This weak-locality holds at cubic level due to Metsaev upper bound and should hold for the quartic self-interactions of the AdS scalar field due to general facts about Mellin amplitudes.

\begin{itemize}
	\item[(a)] \textit{Locality of cubic vertices:} Individual cubic higher-spin interactions are indeed {local} in the sense that 
any 3-point contact Witten diagram with fixed external legs is a polynomial function of the Mellin variables. 
This follows as a corollary from {Metsaev upper bound} (a.ii): For any triplet of spins,
the number of derivatives in any relevant cubic vertex is bounded from above by the sum of the spins. 

	\item[(b)] \textit{Weak-locality of quartic vertices:} The quartic self-interactions of the AdS scalar field dual to the single-trace scalar primary operator in the $O(N)$ model (or, more generally, in a large-$N$ CFT with discrete spectrum) appear to be weakly local, in the sense that the 4-point contact Witten diagram with four scalar external legs is an entire function of the Mellin variables.
 
\vspace{2mm} 
This follows from the following general facts about the decomposition of single-trace scalar primary  operator 4-point function in conformal blocks at leading order in $1/N$:
\begin{itemize}
	\item[(i)] This decomposition contains both single-trace and double-trace conformal blocks. 
	\item[(ii)] Any single-trace conformal block can be accounted by a Witten exchange diagram. 
	\item[(iii)] The Mellin amplitude of a 4-point function {is} an entire function of the Mellin variables iff its conformal block decomposition does \textit{not} contain any single-trace contribution. 
\end{itemize}
The point (i) is expected; the coinciding point limit of two single-trace operators will produce double-trace operators as well.
The point (ii) is AdS/CFT standard lore. The correspondence between single-trace conformal blocks and Witten exchange diagrams is one-to-one: It is essentially an identity up to double-trace conformal blocks (see e.g. the recent discussion in \cite{ElShowk:2011ag}).
The point (iii)	holds because Mellin amplitudes are meromorphic functions with simple poles arising only from single-trace operators \cite{Penedones:2010ue}. 
\end{itemize}
The proof of (C.b) goes as follows: Consider a single-trace scalar primary  operator 4-point function in a CFT with discrete spectrum and at large $N$.
First, decompose this 4-point function into conformal blocks. Second, using facts (i)-(ii) associate to each single-trace conformal block its corresponding Witten exchange diagram. Third, substract from the 4-point function all these Witten exchange diagrams. By construction, the difference that remains is a 4-point function whose decomposition contains only double-trace conformal blocks. Fourth,
conclude from (iii) that this remainder is an entire function of the Mellin variables since it does not contain any single-trace contribution. 
This allows to interpret the remainder as a 4-point contact Witten diagram associated to a weakly local quartic vertex. 
\end{itemize}

As one can see, the boundary criterion of weak locality (see Section \ref{bulkloc}) tantamounts to the existence of a neat separation in the amplitude between the exchange and contact contributions: Poles in the scattering amplitude are accounted by particle exchanges and what remains is interpreted as the contact amplitude. 

\vspace{2mm}
\textbf{Remarks:}
\begin{itemize}
		\item It would be interesting to compare the criterion of weak locality for higher-spin gravity advocated here with the recent proposals \cite{Vasiliev:2015mka} and \cite{Skvortsov:2015lja} 
based, respectively, on functional classes of star-product elements and on classes of field redefinitions leaving Witten diagrams invariant.
		\item A caveat of our argument that the bulk dual of the $O(N)$ model is a weakly local higher-spin gravity is that it implicitly assumes that Mellin amplitudes are well defined functions while they actually require some regularisation. They may also be thought as distributions (c.f. Section \ref{MellinON}) but then it is their analyticity properties which are somewhat elusive. 
		\item  Another subtlety is that the third step in the proof (substraction of all Witten exchange diagrams) can bring infinities in the double-trace conformal block decomposition of the remainder. Taking an optimistic standpoint, this step might actually regularise the Mellin amplitudes alluded above.
\end{itemize}

\section{Holographic reconstruction of higher-spin gravity vertices}\label{quartvertx}

The line of arguments presented in section \ref{HShol} is expected to generalise (to all spins and to higher numbers of points) and indicates that the higher-spin interactions for the bulk dual of the $O(N)$ model might be weakly local. This picture is very suggestive but remains somewhat qualitative. The explicit holographic reconstruction of some quartic vertices in higher-spin gravity, to which we now turn, provides a more concrete playground to test bulk locality.

More specifically, we will consider the holographic reconstruction of quartic AdS interations from a free CFT. In practice, this task amounts to:
\begin{itemize}
		\item Compute the 3 and 4 point conformal correlators via Wick contraction,
		\item Write the most general ansatz for the relevant cubic and quartic vertices,
		\item Compute the corresponding exchange and contact amplitudes, 
		\item Fix the coefficients of vertices by matching correlators with total amplitudes.
\end{itemize}
A priori, it is not guaranteed that such a purely holographic reconstruction produces interactions compatible with the Noether procedure. However, it is natural to expect that these two perturbative procedures are compatible with each other since Ward identities of the boundary CFT should be dual to Noether identities of the AdS theory. 

\begin{figure}[h]
 \centering
\includegraphics[width=0.9\linewidth]{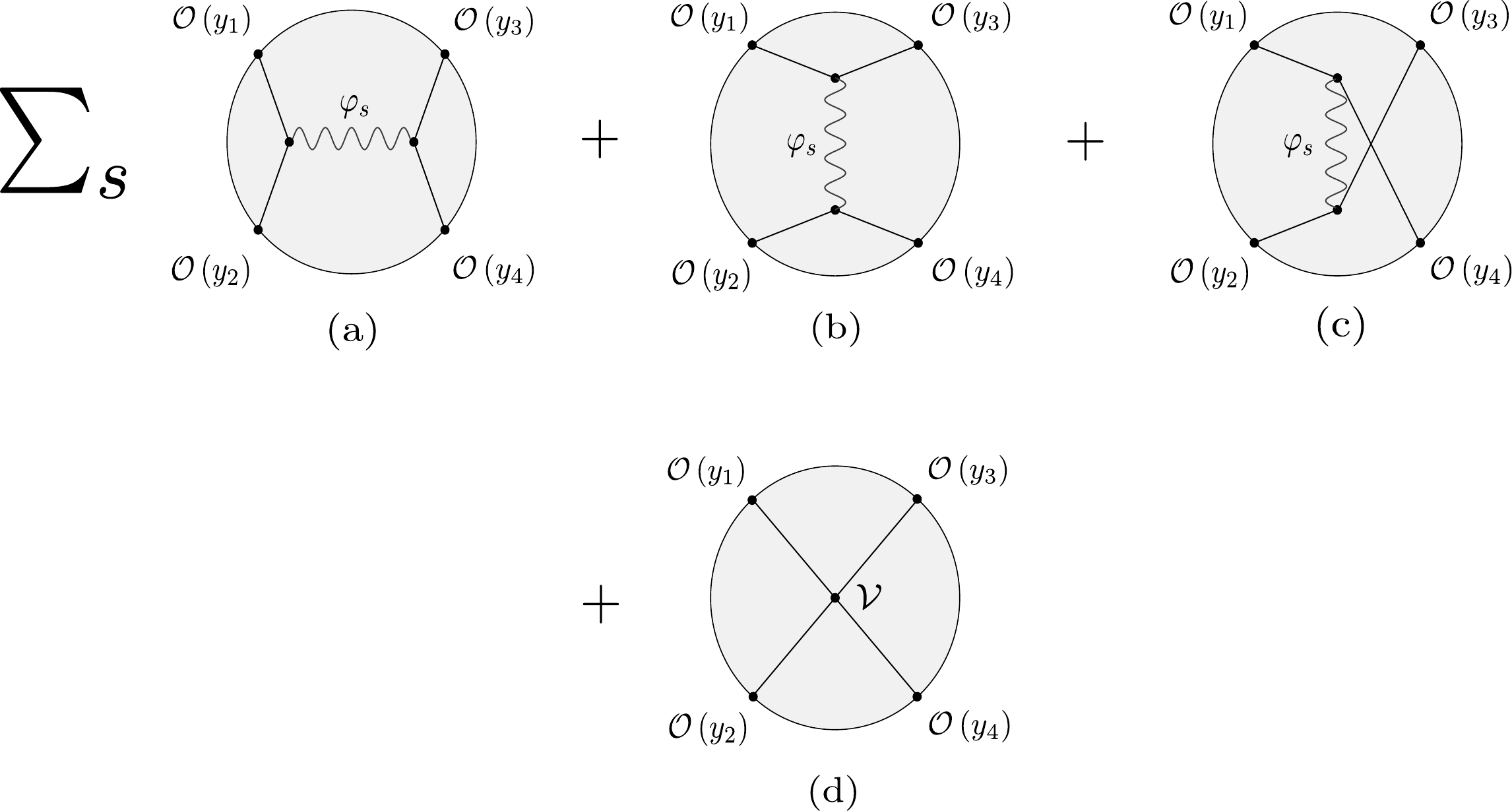}
 \caption{Four-point Witten diagrams contributing to the holographic reconstruction.}
\label{fig::adscft}
 \end{figure}

The holographic reconstruction was performed in \cite{Bekaert:2015tva,Bekaert:2014cea} for the simplest nontrivial case: The quartic self-interaction of the $AdS_4$ scalar field in the higher-spin multiplet dual to the $d=3$ free $O(N)$ model. The relevant four-point Witten diagrams are displayed in figure \ref{fig::adscft}. Diagrams (a)-(c) are exchanges of massless spin-$s$ fields between two pairs of the real scalar. The contact diagram (d) is the amplitude associated to the quartic vertex.

\newpage

\begin{figure}[h]
 \centering
\includegraphics[width=0.7\linewidth]{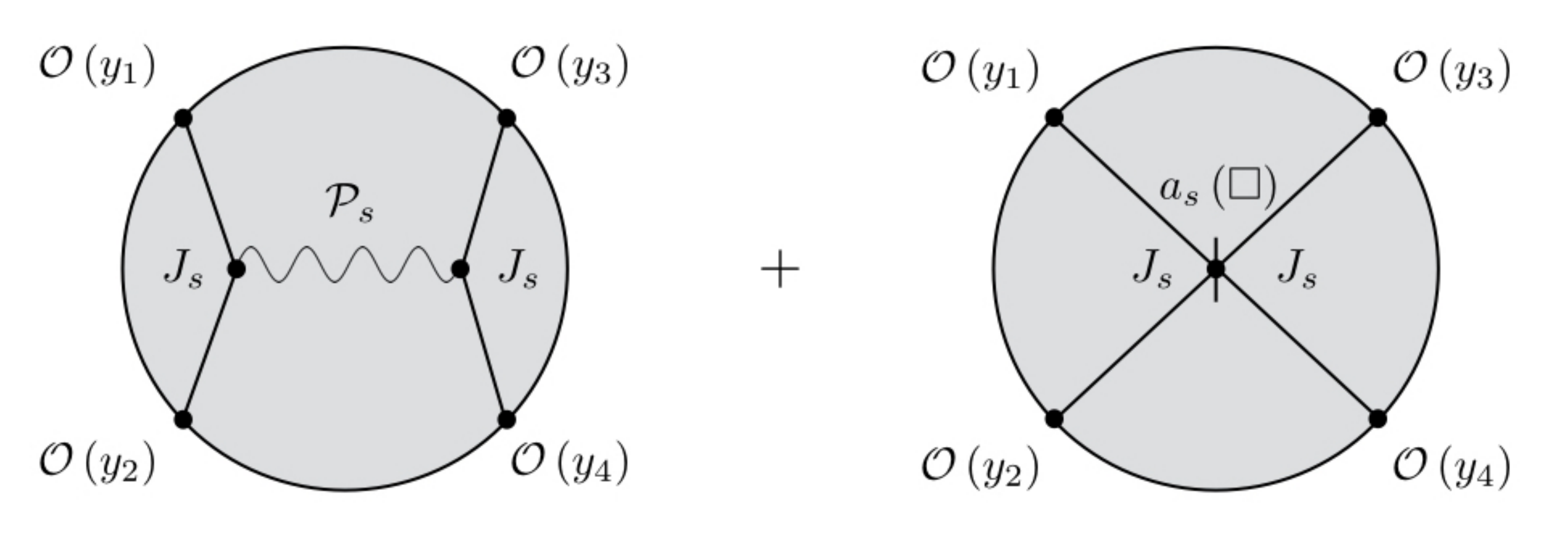}
 \caption{Comparison of exchange and contact Witten diagrams in a single channel.}
\label{fig::singlespin}
 \end{figure}

Among the important technical simplifications in this example are the structure of relevant vertices: 
On the one hand, the bulk cubic vertices $s-0-0$ are of Noether type $\varphi_s J_s$ (see \cite{Berends:1985xx,Bekaert:2010hk}), i.e.
a gauge field $\varphi_s$ times a conserved current $J_s=\varphi_0(\nabla)^s\varphi_0+...$ bilinear in the scalar field $\varphi_0$, and traceless on-shell. Therefore, the diagrams (a)-(c) correspond to current exchange $J_s {\cal P}_s J_s$, where ${\cal P}_s$ is the spin-$s$ gauge field propagator. On the other hand, all bulk quartic vertices $0-0-0-0$ are also of current exchange type $J_s\,\Box^m J_s$ (see \cite{Heemskerk:2009pn,Bekaert:2015tva}). Consequently, the exchange and contact Witten diagrams are of the same type and can be easily compared for each spin $s$ and in each channel (see figure \ref{fig::singlespin}).

The form of the quartic vertex can be summarised as follows:
\begin{equation}
\label{verdef}
\mathcal{V}= \sum\limits_{s\in 2\mathbb{N}} \mathcal{V}_{s}\,,\qquad
\mathcal{V}_{s}\, = \,  J_{\mu_1\cdots \mu_s}\,a_s\left(\Box\right)\,J^{\mu_1\cdots \mu_s},
\end{equation}
where $a_s(\Box)$ are generating functions
\begin{equation}
a_s\left(\Box\right)=\sum\limits_{m=0}^\infty\,a_{m,s}\,\Box^m,
\end{equation}
for the individual couplings $a_{m,s}$ in front of the individual vertices 
\begin{equation}\label{indvert}
\mathcal{V}_{m,s} =J_{\mu_1\cdots \mu_s}\,\Box^m\,J^{\mu_1\cdots \mu_s}\,. 
\end{equation}
The explicit expression of the quartic vertex $\mathcal{V}$ can be found in \cite{Bekaert:2015tva}.
The generating functions are defined by
\begin{align}\nonumber
&a_s\left(\nu^2+s+\tfrac{9}{4}\right)
\propto\frac{2^{8-s}}{\nu ^2+(s-\tfrac{1}{2})^2}\left[\frac{\pi}{\Gamma \left(\frac{2 s-2 i \nu +1}{4}\right)^2 \Gamma \left(\frac{2 s+2 i \nu +1}{4} \right)^2} - \frac{1}{ \Gamma\left(s\right)^2}\right] \\ \label{answervertex}
 &\, - \frac{ \left(-1\right)^{\tfrac{s}{2}} \pi ^{\frac{3}{2}} 2^{s+5} \Gamma \left(s+\frac{3}{2}\right)\Gamma \left(\frac{s}{2}+\frac{1}{2}\right)}{\sqrt{2}\,\Gamma \left(\frac{s}{2}+1\right)\Gamma\left(s+1\right)\Gamma \left(\frac{3}{4}-\frac{i \nu }{2}\right) \Gamma \left(\frac{3}{4}+\frac{i \nu }{2}\right) \Gamma \left(s+\tfrac{1}{2}+ i \nu \right) \Gamma \left(s+\tfrac{1}{2}- i \nu \right)}\,,\nonumber
\end{align}
which are entire functions, though it is not manifest. Heuristically, this indicates that each contact diagram associated to the cubic vertex $\mathcal{V}_{s}$ and depicted in figure \ref{fig::singlespin} looks weakly local. 

A subtlety is that the quartic vertices
\eqref{indvert} are linearly independent on the free mass shell only for $2m\leqslant s$. However, the generating function includes infinitely many terms with $2m>s$. The latter should be reexpressed in terms of the previous independent ones. Since this involves infinitely many couplings and since we are on $AdS_4$ background where covariant derivatives do not commute, this procedure can a priori produce infinities, c.f. the examples in \cite{Boulanger:2015ova,Skvortsov:2015lja} at cubic level (see also \cite{Giombi:2010vg,
Boulanger:2008tg,Bekaert:2010hw
} where this possibility was discussed). In other words, the couplings in front of the independent quartic vertices might be infinite. 
However, one may expect this phenomenon not to happen in the present context of holographic reconstruction because 
infinite couplings in front of the independent quartic vertices would correspond to infinite coefficients in the double-trace conformal block decomposition.

\section{Mellin amplitudes of the free and critical $O(N)$ model}\label{MellinON}

As mentioned above, Mellin amplitudes of weakly-coupled theories are sometimes ill-defined as genuine functions, so they may either require some regularisation procedure or to be treated as generalised functions (i.e. distributions). We will investigate the latter option and focus on the $O(N)$ model.

The scalar single-trace primary operator $\cal O$ has conformal dimension $\Delta=d-2$ for the $d$-dimensional free $O(N)$ model.
Its 4-point correlator \eqref{4pt} has a factor not fixed by conformal symmetry
\begin{equation}
F_{\mbox{free}}(u,v)=F_{\mbox{dis.}}(u,v)+\frac1{N}\,F_{\mbox{con.}}(u,v) 
\end{equation}
where the first term $F_{\mbox{dis.}}(u,v)$ arise from disconnected diagrams and the second term $F_{\mbox{con.}}(u,v)$ from connected diagrams.
Both terms are power functions of the two cross ratios. This is true for the disconnected piece
\begin{equation}\label{Fdisc}
F_{\mbox{dis.}}(u,v)= 1+u^\Delta+\left(\frac{u}{v}\right)^\Delta
\end{equation}
and for the connected piece
\begin{equation}\label{Fconn}
F_{\mbox{con.}}(u,v)= u^{\frac{\Delta}{2}}+\left(\frac{u}{v}\right)^{\frac{\Delta}{2}}+u^{\frac{\Delta}{2}}\left(\frac{u}{v}\right)^{\frac{\Delta}{2}}\,,
\end{equation}
where, again, front numerical factors have been neglected for the sake of simplicity.

A power function, say 
\begin{equation}\label{Mtransfo}
f(x)=x^\alpha\qquad (\alpha\in{\mathbb R})\,,
\end{equation}
does not have a well defined Mellin transform \cite{Flajolet19953}
\begin{equation}\label{Mtransfoo}
M(z)=\int\limits_0^\infty x^z\,f(x)\,\frac{dx}{x}\,,
\end{equation}
or, rather, its fundamental strip of convergence degenerates to the vertical line $z=-\,\alpha+i\,\mathbb R$ in the complex plane. Actually, the Mellin transform \eqref{Mtransfoo} of the power function \eqref{Mtransfo} may be understood$^*$\let\thefootnote\relax\footnote{$^*$ We thank S.~Rychkov and M.~A.~Vasiliev for pointing out to us this possibility, which we later traced in the mathematical litterature.}
as a distribution \cite{Mellintransform}:
$M(z)=\delta(z+\alpha)$\,.
With this interpretation, the Mack ampitudes corresponding to \eqref{Fdisc} and \eqref{Fconn} are, respectively, 
\begin{equation}\label{Mack1}
M_{\mbox{dis.}}(s,t)= \delta(s)\,\delta(t-\Delta)+\delta(s-\Delta)\,\delta(t)+\delta(s-\Delta)\,\delta(t-\Delta)
\end{equation}
and 
\begin{equation}\label{Mack2}
M_{\mbox{con.}}(s,t)= \delta(s-\tfrac{\Delta}2)\,\delta(t-\tfrac{\Delta}2)+\delta(s-\tfrac{\Delta}2)\,\delta(t-\Delta)+\delta(s-\Delta)\,\delta(t-\tfrac{\Delta}2)\,.
\end{equation}
Due to the distributional nature of these reduced Mellin amplitudes $M_{\mbox{dis.}}(s,t)$ and $M_{\mbox{con.}}(s,t)$, the definition  \eqref{Mst} for the corresponding Mellin amplitudes ${\cal M}_{\mbox{dis.}}(s,t)$ and ${\cal M}_{\mbox{con.}}(s,t)$ is difficult to apply. The problem is that each value in the support of the left-hand-side in \eqref{Mst} corresponds to a pole at zero of one (even two for the disconnected piece) gamma function on the right-hand-side. Formally, treating the Mellin amplitudes ${\cal M}_{\mbox{dis.}}(s,t)$ and ${\cal M}_{\mbox{con.}}(s,t)$ as genuine functions, one would conclude that they must vanish identically. In other words, although the reduced Mellin amplitudes \eqref{Mack1} and 
\eqref{Mack2} are well defined distributions, the nature of the corresponding Mellin amplitudes remains to be clarified.

Let us stress that the above observations are not restricted to free CFTs but even holds for some integrable CFTs. A good example is the large-$N$ limit of the $O(N)$ model at large $N$ whose integrability has been recognised a long time ago \cite{Stanley:1968gx}. From a modern perspective, the integrability of the large-$N$ limit of the $O(N)$ model is intimately related to its equivalence to the free $O(N)$ model, up to a Legendre transformation. This latter fact lies at the heart of the Klebanov-Polyakov conjecture \cite{Klebanov:2002ja}. In the $N\to\infty$ limit, the scalar single-trace primary operator $\cal O$ for the critical $O(N)$ model has conformal dimension $\Delta=2$ and its 4-point correlator \eqref{4pt} has a factor not fixed by conformal symmetry
\begin{equation}
F_{\mbox{int}}(u,v)=G_{\mbox{dis.}}(u,v)+\frac1{N}\,G_{\mbox{con.}}(u,v)+\,O\Big(\frac1{N^2}\Big) 
\end{equation}
which is also a sum of power functions of the two cross ratios. More precisely, the disconnected part is given by formula \eqref{Fdisc} with $\Delta=2$ and the connected part reads for $d=3$ \cite{Leonhardt:2003du}
\begin{equation}\label{Fconnint}
G_{\mbox{con.}}(u,v)= u^2\left[u^{-\frac{3}{2}}\left(1+u-v\right)+v^{-\frac{3}{2}}\left(1-u+v\right)-(uv)^{-\frac{3}{2}}\left(1-u-v\right)\right]\,.
\end{equation}
The reduced Mellin amplitude of the connected part is thus
\begin{align}
M_{\mbox{con.}}(s,t) &= \delta(s-\tfrac12)\,\delta(t-\tfrac32)+\delta(s-\tfrac32)\,\delta(t-\tfrac12)-\delta(s-\tfrac12)\,\delta(t-\tfrac12)\nonumber\\
&+\,\delta(s-2)\,\delta(t-\tfrac32)-\delta(s-3)\,\delta(t-\tfrac12)+\delta(s-2)\,\delta(t-\tfrac12)\label{Mack3}\\
&-\,\delta(s-\tfrac12)\,\delta(t-3)+\delta(s-\tfrac32)\,\delta(t-2)+\delta(s-\tfrac12)\,\delta(t-2)\,.\nonumber
\end{align}
Again the support of the Dirac delta functions coincide with poles of the gamma functions in \eqref{Mst}.
The presence of a product of two Dirac distributions in the reduced Mellin amplitudes \eqref{Mack1}, \eqref{Mack2} and \eqref{Mack3} seem to preclude their interpretation as scattering amplitudes since trivial scattering (i.e. alignment of pairs of momenta) should correspond to fixing a single Mandelstam variable (and not all of them). This may be a hint that Mellin amplitudes of free (or integrable) CFTs do not admit a sensible flat limit, a reasonable expectation since unbroken higher-spin gravity theories do not admit a weakly-coupled flat regime.

The distributional nature of tree-level amplitudes of four scalar fields and their absence of scattering has also been recently observed in conformal higher-spin gravity \cite{Joung:2015eny}. The triviality of scattering amplitudes may look surprising at first sight since higher-spin gravity theory has nonvanishing bulk interaction vertices. Nevertheless, retrospectively it seems reasonable to expect that the algebra of asymptotic higher-spin symmetries is so huge as to constrain the scattering of particles to be essentially trivial.

\section{Conclusion and open directions}

Characterising the degree of (non)locality of the bulk duals of weakly-coupled CFTs is an important issue for understanding holographic duality beyond the regime where the bulk spacetime is weakly curved.
Looking at higher-spin gravity suggests to broaden the definition of locality 
in order to encompass interactions on flat/AdS spacetime for which the Feynman/Witten amplitudes of the corresponding contact diagrams are not polynomial but entire functions. At the level of four-point conformal correlators and quartic bulk vertices, this weak locality criterion is equivalent to the property that Witten exchange diagrams account for all single-trace conformal blocks in the decomposition of correlators. General arguments and explicit computations indicate that, for the higher-spin gravity dual of the free $O(N)$ model, the quartic self-interaction of the bulk scalar field is weakly (non)local.

Extending the holographic reconstruction of quartic vertices to spin $s\neq 0$ using twistor techniques, or to boundary dimension $d\neq 3$, are interesting challenges, as well as comparing the quartic vertices obtained purely from holographic reconstruction with the corresponding ones that can be extracted from Vasiliev equations.


\section*{Acknowledgments}

We thank E.~Joung, E.~Skvortsov, M.~Taronna and M.~A.~Vasiliev for useful discussions.

X.B. is very grateful to the Institute of Advanced Studies from Nanyang Technological University in Singapore for hospitality, and to the organisers of the workshop ``Higher Spin Gauge Theories'' for providing the opportunity to present this work and contribute to the proceedings.
He also acknowledges the Asia Pacific Center for Theoretical Physics (APCTP) in Pohang
and the Institute for Studies in Theoretical Physics and Mathematics (IPM) in Tehran
for hospitality where part of this work was done during, respectively, the program ``Duality and Novel Geometry in M-theory''
and the ``IPM school on Higher Spin Theory''.
D.P. thanks E. Joung and Seoul National University for their kind hospitality.

The research of X.B. was supported by the Russian Science Foundation grant 14-42-00047 in association with Lebedev Physical Institute.
The work of J.E. and C.S. was partially supported by the European Science Foundation Holograv network (Holographic methods for strongly coupled systems). The work of D.P. was supported by the ERC Advanced grant No.290456.

\providecommand{\href}[2]{#2}\begingroup\raggedright\endgroup


\begin{thebibliography}{10}

\bibitem{Heemskerk:2009pn}
I.~Heemskerk, J.~Penedones, J.~Polchinski, and J.~Sully, ``{Holography from
  Conformal Field Theory},''
  \href{http://dx.doi.org/10.1088/1126-6708/2009/10/079}{{\em JHEP} {\bfseries
  0910} (2009) 079},
\href{http://arxiv.org/abs/0907.0151}{{\ttfamily arXiv:0907.0151 [hep-th]}}.

\bibitem{Fitzpatrick:2010zm}
A.~L. Fitzpatrick, E.~Katz, D.~Poland, and D.~Simmons-Duffin, ``{Effective
  Conformal Theory and the Flat-Space Limit of AdS},''
  \href{http://dx.doi.org/10.1007/JHEP07(2011)023}{{\em JHEP} {\bfseries 07}
  (2011) 023},
\href{http://arxiv.org/abs/1007.2412}{{\ttfamily arXiv:1007.2412 [hep-th]}}.

\bibitem{Fitzpatrick:2012cg}
A.~L. Fitzpatrick and J.~Kaplan, ``{AdS Field Theory from Conformal Field
  Theory},'' \href{http://dx.doi.org/10.1007/JHEP02(2013)054}{{\em JHEP}
  {\bfseries 02} (2013) 054},
\href{http://arxiv.org/abs/1208.0337}{{\ttfamily arXiv:1208.0337 [hep-th]}}.

\bibitem{Penedones:2010ue}
J.~Penedones, ``{Writing CFT correlation functions as AdS scattering
  amplitudes},'' \href{http://dx.doi.org/10.1007/JHEP03(2011)025}{{\em JHEP}
  {\bfseries 03} (2011) 025},
\href{http://arxiv.org/abs/1011.1485}{{\ttfamily arXiv:1011.1485 [hep-th]}}.

\bibitem{Mack:2009mi}
G.~Mack, ``{D-independent representation of Conformal Field Theories in D
  dimensions via transformation to auxiliary Dual Resonance Models. Scalar
  amplitudes},''
\href{http://arxiv.org/abs/0907.2407}{{\ttfamily arXiv:0907.2407 [hep-th]}}.

\bibitem{Paulos:2011ie}
M.~F. Paulos, ``{Towards Feynman rules for Mellin amplitudes},''
  \href{http://dx.doi.org/10.1007/JHEP10(2011)074}{{\em JHEP} {\bfseries 10}
  (2011) 074},
\href{http://arxiv.org/abs/1107.1504}{{\ttfamily arXiv:1107.1504 [hep-th]}}.

\bibitem{Fitzpatrick:2011ia}
A.~L. Fitzpatrick, J.~Kaplan, J.~Penedones, S.~Raju, and B.~C. van Rees, ``{A
  Natural Language for AdS/CFT Correlators},''
  \href{http://dx.doi.org/10.1007/JHEP11(2011)095}{{\em JHEP} {\bfseries 11}
  (2011) 095},
\href{http://arxiv.org/abs/1107.1499}{{\ttfamily arXiv:1107.1499 [hep-th]}}.

\bibitem{ElShowk:2011ag}
S.~El-Showk and K.~Papadodimas, ``{Emergent Spacetime and Holographic CFTs},''
  \href{http://dx.doi.org/10.1007/JHEP10(2012)106}{{\em JHEP} {\bfseries 1210}
  (2012) 106},
\href{http://arxiv.org/abs/1101.4163}{{\ttfamily arXiv:1101.4163 [hep-th]}}.

\bibitem{Coleman:1967ad}
S.~R. Coleman and J.~Mandula, ``{All Possible Symmetries of the S Matrix},''
\href{http://dx.doi.org/10.1103/PhysRev.159.1251}{{\em Phys. Rev.} {\bfseries
  159} (1967) 1251--1256}.

\bibitem{Bekaert:2015tva}
X.~Bekaert, J.~Erdmenger, D.~Ponomarev, and C.~Sleight, ``{Quartic AdS
  Interactions in Higher-Spin Gravity from Conformal Field Theory},''
  \href{http://dx.doi.org/10.1007/JHEP11(2015)149}{{\em JHEP} {\bfseries 11}
  (2015) 149},
\href{http://arxiv.org/abs/1508.04292}{{\ttfamily arXiv:1508.04292 [hep-th]}}.

\bibitem{Sundborg:2000wp}
B.~Sundborg, ``{Stringy gravity, interacting tensionless strings and massless
  higher spins},'' \href{http://dx.doi.org/10.1016/S0920-5632(01)01545-6}{{\em
  Nucl. Phys. Proc. Suppl.} {\bfseries 102} (2001) 113--119},
  \href{http://arxiv.org/abs/hep-th/0103247}{{\ttfamily arXiv:hep-th/0103247
  [hep-th]}}.
[,113(2000)];\\
S.~E. Konstein, M.~A. Vasiliev, and V.~N. Zaikin, ``{Conformal higher spin
  currents in any dimension and AdS / CFT correspondence},''
  \href{http://dx.doi.org/10.1088/1126-6708/2000/12/018}{{\em JHEP} {\bfseries
  12} (2000) 018},
\href{http://arxiv.org/abs/hep-th/0010239}{{\ttfamily arXiv:hep-th/0010239
  [hep-th]}};\\
E.~Sezgin and P.~Sundell, ``{Doubletons and 5-D higher spin gauge theory},''
  \href{http://dx.doi.org/10.1088/1126-6708/2001/09/036}{{\em JHEP} {\bfseries
  09} (2001) 036},
\href{http://arxiv.org/abs/hep-th/0105001}{{\ttfamily arXiv:hep-th/0105001
  [hep-th]}};\\
A.~Mikhailov, ``{Notes on higher spin symmetries},''
\href{http://arxiv.org/abs/hep-th/0201019}{{\ttfamily arXiv:hep-th/0201019
  [hep-th]}};\\
E.~Sezgin and P.~Sundell, ``{Massless higher spins and holography},''
  \href{http://dx.doi.org/10.1016/S0550-3213(02)00739-3}{{\em Nucl. Phys.}
  {\bfseries B644} (2002) 303--370},
  \href{http://arxiv.org/abs/hep-th/0205131}{{\ttfamily arXiv:hep-th/0205131
  [hep-th]}}.
[Erratum: Nucl. Phys.B660,403(2003)].

\bibitem{Klebanov:2002ja}
I.~R. Klebanov and A.~M. Polyakov, ``{AdS dual of the critical O(N) vector
  model},'' \href{http://dx.doi.org/10.1016/S0370-2693(02)02980-5}{{\em Phys.
  Lett.} {\bfseries B550} (2002) 213--219},
\href{http://arxiv.org/abs/hep-th/0210114}{{\ttfamily arXiv:hep-th/0210114
  [hep-th]}}.

\bibitem{Sezgin:2003pt}
E.~Sezgin and P.~Sundell, ``{Holography in 4D (super) higher spin theories and
  a test via cubic scalar couplings},''
  \href{http://dx.doi.org/10.1088/1126-6708/2005/07/044}{{\em JHEP} {\bfseries
  0507} (2005) 044},
\href{http://arxiv.org/abs/hep-th/0305040}{{\ttfamily arXiv:hep-th/0305040
  [hep-th]}}.

\bibitem{Metsaev:2005ar}
R.~R. Metsaev, ``{Cubic interaction vertices of massive and massless higher
  spin fields},'' \href{http://dx.doi.org/10.1016/j.nuclphysb.2006.10.002}{{\em
  Nucl. Phys.} {\bfseries B759} (2006) 147--201},
\href{http://arxiv.org/abs/hep-th/0512342}{{\ttfamily arXiv:hep-th/0512342
  [hep-th]}}.

\bibitem{Fradkin:1986ka}
E.~S. Fradkin and M.~A. Vasiliev, ``{Candidate to the Role of Higher Spin
  Symmetry},''
\href{http://dx.doi.org/10.1016/S0003-4916(87)80025-8}{{\em Annals Phys.}
  {\bfseries 177} (1987) 63}.

\bibitem{Boulanger:2013zza}
N.~Boulanger, D.~Ponomarev, E.~Skvortsov, and M.~Taronna, ``{On the uniqueness
  of higher-spin symmetries in AdS and CFT},''
  \href{http://dx.doi.org/10.1142/S0217751X13501625}{{\em Int.J.Mod.Phys.}
  {\bfseries A28} (2013) 1350162},
\href{http://arxiv.org/abs/1305.5180}{{\ttfamily arXiv:1305.5180 [hep-th]}}.

\bibitem{Vasiliev:2015mka}
M.~A. Vasiliev, ``{Star-Product Functions in Higher-Spin Theory and
                        Locality},''
  \href{http://dx.doi.org/10.1007/JHEP06(2015)031}{{\em JHEP} {\bfseries 06}
  (2015) 031},
\href{http://arxiv.org/abs/1502.02271}{{\ttfamily arXiv:1502.02271 [hep-th]}}.

\bibitem{Skvortsov:2015lja}
E.~D. Skvortsov and M.~Taronna, ``{On Locality, Holography and Unfolding},''
  \href{http://dx.doi.org/10.1007/JHEP11(2015)044}{{\em JHEP} {\bfseries 11}
  (2015) 044},
\href{http://arxiv.org/abs/1508.04764}{{\ttfamily arXiv:1508.04764 [hep-th]}}.

\bibitem{Bekaert:2014cea}
X.~Bekaert, J.~Erdmenger, D.~Ponomarev, and C.~Sleight, ``{Towards holographic higher-spin interactions: Four-point
                        functions and higher-spin exchange},''
  \href{http://dx.doi.org/10.1007/JHEP03(2015)170}{{\em JHEP} {\bfseries 03}
  (2015) 170},
\href{http://arxiv.org/abs/1412.0016}{{\ttfamily arXiv:1412.0016 [hep-th]}}.

\bibitem{Berends:1985xx}
F.~A. Berends, G.~J.~H. Burgers, and H.~van Dam, ``{Explicit Construction of
  Conserved Currents for Massless Fields of Arbitrary Spin},''
{\em Nucl. Phys.} {\bfseries B271} (1986) 429.

\bibitem{Bekaert:2010hk}
X.~Bekaert and E.~Meunier, ``{Higher spin interactions with scalar matter on
  constant curvature spacetimes: conserved current and cubic coupling
  generating functions},''
  \href{http://dx.doi.org/10.1007/JHEP11(2010)116}{{\em JHEP} {\bfseries 1011}
  (2010) 116},
\href{http://arxiv.org/abs/1007.4384}{{\ttfamily arXiv:1007.4384 [hep-th]}}.

\bibitem{Boulanger:2015ova}
N.~Boulanger, P.~Kessel, E.~D. Skvortsov, and M.~Taronna, ``{Higher Spin
  Interactions in Four Dimensions: Vasiliev vs. Fronsdal},''
  \href{http://dx.doi.org/10.1088/1751-8113/49/9/095402}{{\em J. Phys.}
  {\bfseries A49} no.~9, (2016) 095402},
\href{http://arxiv.org/abs/1508.04139}{{\ttfamily arXiv:1508.04139 [hep-th]}}.

\bibitem{Giombi:2010vg}
S.~Giombi and X.~Yin, ``{Higher Spins in AdS and Twistorial Holography},''
  \href{http://dx.doi.org/10.1007/JHEP04(2011)086}{{\em JHEP} {\bfseries 04}
  (2011) 086},
\href{http://arxiv.org/abs/1004.3736}{{\ttfamily arXiv:1004.3736 [hep-th]}};
``{Higher Spin Gauge Theory and Holography: The
  Three-Point Functions},''
  \href{http://dx.doi.org/10.1007/JHEP09(2010)115}{{\em JHEP} {\bfseries 09}
  (2010) 115},
\href{http://arxiv.org/abs/0912.3462}{{\ttfamily arXiv:0912.3462 [hep-th]}}.

\bibitem{Boulanger:2008tg}
N.~Boulanger, S.~Leclercq, and P.~Sundell, ``{On The Uniqueness of Minimal
  Coupling in Higher-Spin Gauge Theory},''
  \href{http://dx.doi.org/10.1088/1126-6708/2008/08/056}{{\em JHEP} {\bfseries
  08} (2008) 056},
\href{http://arxiv.org/abs/0805.2764}{{\ttfamily arXiv:0805.2764 [hep-th]}}.

\bibitem{Bekaert:2010hw}
X.~Bekaert, N.~Boulanger, and P.~Sundell, ``{How higher-spin gravity surpasses
  the spin two barrier: no-go theorems versus yes-go examples},''
  \href{http://dx.doi.org/10.1103/RevModPhys.84.987}{{\em Rev. Mod. Phys.}
  {\bfseries 84} (2012) 987--1009},
\href{http://arxiv.org/abs/1007.0435}{{\ttfamily arXiv:1007.0435 [hep-th]}}.

\bibitem{Flajolet19953}
P.~Flajolet, X.~Gourdon, and P.~Dumas, ``Mellin transforms and asymptotics:
  Harmonic sums,''
  \href{http://dx.doi.org/http://dx.doi.org/10.1016/0304-3975(95)00002-E}{{\em
  Theor. Comp. Sc.} {\bfseries 144} (1995) 3}.

\bibitem{Mellintransform}
J.~Bertrand, P.~Bertrand, and J.-P. Ovarlez, ``The Mellin transform,'' in A.~D. Poularikas (ed.), {\em
  Transforms and applications handbook} (CRC Press, 2000) chapter 11.

\bibitem{Stanley:1968gx}
H.~E. Stanley, ``{Spherical model as the limit of infinite spin
  dimensionality},''
\href{http://dx.doi.org/10.1103/PhysRev.176.718}{{\em Phys. Rev.} {\bfseries
  176} (1968) 718--722}.

\bibitem{Leonhardt:2003du}
T.~Leonhardt and W.~R\"uhl, ``{The minimal conformal O(N) vector sigma model at d
  = 3},'' \href{http://dx.doi.org/10.1088/0305-4470/37/4/023}{{\em J. Phys.}
  {\bfseries A37} (2004) 1403--1413},
\href{http://arxiv.org/abs/hep-th/0308111}{{\ttfamily arXiv:hep-th/0308111
  [hep-th]}}.

\bibitem{Joung:2015eny}
E.~Joung, S.~Nakach, and A.~A. Tseytlin, ``{Scalar scattering via conformal
  higher spin exchange},''
  \href{http://dx.doi.org/10.1007/JHEP02(2016)125}{{\em JHEP} {\bfseries 02}
  (2016) 125},
\href{http://arxiv.org/abs/1512.08896}{{\ttfamily arXiv:1512.08896 [hep-th]}}.

\end{thebibliography}
\end{document}